\documentclass[10pt,conference]{IEEEtran}
  \IEEEoverridecommandlockouts
  
\usepackage{cite}
\usepackage[symbol]{footmisc}

\usepackage{graphicx}
{\fnsymbol{footnote}}
\usepackage{amsmath,amssymb,amsfonts}
\usepackage{algorithmic}
\usepackage{textcomp}
\usepackage{xcolor}
\IEEEoverridecommandlockouts
\usepackage[linesnumbered,ruled,vlined]{algorithm2e}
\usepackage{subcaption}
\usepackage{epstopdf}
\usepackage{setspace}
\usepackage{multicol}
\usepackage{color}
\usepackage{romannum}
\usepackage{tabularx}
\usepackage{booktabs}
\usepackage{caption}
\usepackage{url} 
\usepackage[left=0.625in,right=0.625in,top=0.75in,bottom=1in]{geometry}
\usepackage[linesnumbered,ruled,vlined]{algorithm2e}
\newcommand*{\rom}[1]{\expandafter\@slowromancap\romannumeral #1@}
\def\BibTeX{{\rm B\kern-.05em{\sc i\kern-.025em b}\kern-.08em
    T\kern-.1667em\lower.7ex\hbox{E}\kern-.125emX}}
\begin{document}

\title{PQLS: A High-Performance Python Library for Steady-State Simulation of Open Quantum Systems}
\author{
\IEEEauthorblockN{
Evan Simanovskis\IEEEauthorrefmark{1},
Raviraj Adve\IEEEauthorrefmark{2},
and Javane Rostampoor\IEEEauthorrefmark{1}\IEEEauthorrefmark{2}
}

\IEEEauthorblockA{
\IEEEauthorrefmark{1}
\textit{Dept. of Computer and Mathematical Sciences},
\textit{University of Toronto Scarborough},
Toronto, Canada
}

\IEEEauthorblockA{
\IEEEauthorrefmark{2}
\textit{Dept. of Electrical and Computer Engineering},
\textit{University of Toronto},
Toronto, Canada
}

\IEEEauthorblockA{
Email: evan.simanovskis@mail.utoronto.ca,
rsadve@ece.utoronto.ca,
javane.rostampoor@utoronto.ca
}
}




\maketitle
\begin{abstract}

PQLS (Parallel Quantum Liouvillian Solver) is a high-performance Python
library for computing steady-state solutions of the Lindblad master equation.
It provides a layered user-facing API with three levels of abstraction. The
high-level API accepts physical descriptions of atomic ladder systems and
automatically determines the required model parameters; the mid-level API
accepts quantities such as Rabi frequencies, detunings, decay rates, and
network topology; finally the low-level API allows users to directly specify the
Hamiltonian and collapse operators. This layered design enables users to choose
between physical convenience and direct numerical control depending on the
application.
PQLS is built on JAX and XLA to enable vectorized and hardware-accelerated
computation of large parameter sweeps. Rather than solving the steady-state
problem sequentially for each parameter configuration, PQLS represents
multiple Hamiltonians as batched tensors and processes them through a 
JAX-compiled solver, reducing Python-level overhead and improving hardware
utilization. This is particularly useful for applications requiring large or
multidimensional parameter sweeps, such as computing frequency spectra,
evaluating field-dependent responses, and performing Doppler averaging over
atomic velocity distributions. PQLS has been benchmarked against QuTiP,
QuTiP-JAX, and RydIQule on CPU and GPU platforms, demonstrating substantial
computational speedups across the tested configurations.

\end{abstract}

\begin{IEEEkeywords}
Lindblad master equation, Open quantum systems, Steady-state simulation, JAX, Hardware acceleration
\end{IEEEkeywords}
\section{Introduction}

Open quantum systems arise when a quantum system interacts with its surrounding
environment, leading to effects such as spontaneous emission, dephasing, and
population decay~\cite{agarwal2012open}. Their behavior is commonly modeled using the Lindblad master
equation, in which the Hamiltonian describes the coherent dynamics of the
system, including the energy-level structure, detunings, and driven couplings,
while collapse operators represent dissipative processes~\cite{linblad2}. These coherent and dissipative contributions can be combined into a
Liouvillian superoperator, which governs the evolution of the
system's quantum state as described by its density matrix~\cite{Liouvillan}.
In many applications,
the primary quantity of interest is the steady-state density matrix, from which
state populations, quantum coherences, and experimentally relevant observables
such as optical absorption and transmission can be obtained.

When an electromagnetic field resonantly drives a transition between two
quantum states, the population oscillates coherently between them at the Rabi
frequency, which is determined by the transition dipole moment and the applied
field amplitude. Rydberg atoms, whose highly excited states possess large
electric-dipole moments, are therefore particularly sensitive to electromagnetic
fields and are widely used in atomic sensing and spectroscopy~\cite{Rabi, noise}. 

The field-sensitive interactions give rise to spectroscopic phenomena such as
electromagnetically induced transparency (EIT) and Autler--Townes splitting
(ATS). In EIT, coherent interference between driven transitions suppresses
absorption over a narrow frequency range~\cite{EIT}. In ATS, an additional
external electromagnetic field couples two states and forms two dressed states,
producing a characteristic splitting of the spectral feature~\cite{decay2}. These effects are
typically analyzed in the frequency domain by sweeping one or more detunings,
i.e., the difference between an applied driving frequency and the corresponding
transition resonance. Each detuning value defines a different system
configuration and therefore requires an independent steady-state solution.
Consequently, constructing a single spectrum may require solving hundreds or
thousands of independent steady-state problems across closely related parameter
configurations.


The computational burden becomes substantially larger when additional physical
effects are included. For example, Doppler averaging requires evaluating the
optical response over a distribution of atomic velocities for every detuning
value~\cite{Doppler}. Similar computational complexity is present when sweeping Rabi frequencies,
electric-field amplitudes, decay rates, temperatures, or combinations of these
parameters. Such multidimensional parameter sweeps can therefore require the
solution of very large numbers of independent steady-state problems and may
become the dominant computational cost in numerical studies of open quantum
systems.

Several established software packages provide tools to model and solve
open quantum systems. QuTiP is a widely used general-purpose framework for
constructing quantum models and solving master equations, while QuTiP-JAX
introduces JAX-backed numerical data structures and operations~\cite{qutip5, qutipjax}. RydIQule
provides higher-level capabilities targeted toward atomic and Rydberg sensing
applications, including spectroscopy and Doppler-aware simulations~\cite{rydiqule2024}. In
addition, the Alkali Rydberg Calculator (ARC) provides atomic-structure
information such as transition frequencies, dipole matrix elements, and state
lifetimes~\cite{arc2017}. These tools address complementary aspects of quantum-system
simulation; however, large parameter sweeps can still incur substantial
computational overhead when independent system configurations are constructed
and solved sequentially.

To address this issue, we introduce PQLS (Parallel Quantum Liouvillian Solver)%
\footnote{PQLS is available as open-source software at
\url{https://github.com/Semyazi/pqls} and as an installable package on PyPI at
\url{https://pypi.org/project/pqls/}.},
a high-performance Python library for steady-state simulation of open quantum
systems. PQLS is designed to treat collections of independent parameter
configurations as batched numerical problems rather than as repeated sequential
solver calls. Built on JAX and XLA, it supports vectorized execution and
hardware acceleration on CPUs and supported GPUs, reducing Python-level
overhead and improving utilization of computational resources.

PQLS provides a layered Application Programming Interface (API), 
comprising
the user-facing functions and data structures through which a quantum
system is specified and solved. The high-level API accepts physical
descriptions of atomic ladder systems and uses ARC to determine the required
atomic parameters, including transition frequencies, dipole matrix elements,
Rabi frequencies, and decay rates. The intermediate API accepts quantities such
as Rabi frequencies, detunings, decay rates, and arbitrary network topologies,
while the low-level API allows users to directly provide the Hamiltonian and
collapse operators. This structure enables users to choose between a
physics-oriented description of the system and direct numerical control,
depending on the application.

The main contributions of PQLS are:
\begin{itemize}
    \item a JAX-compiled steady-state solver designed for vectorized and batched
    parameter sweeps;
    \item a three-tier API supporting physical
    atomic ladder models, arbitrary quantum-network topologies, and direct
    Hamiltonian and collapse-operator input;
    \item integration with ARC for automated construction of atomic parameters
    in high-level simulations;
    \item hardware-independent execution on CPUs and supported GPUs through
    JAX/XLA; and
    \item numerical validation and performance benchmarking against QuTiP,
    QuTiP-JAX, and RydIQule, demonstrating substantial speedups for batched
steady-state parameter sweeps.
\end{itemize}

The remainder of this paper is organized as follows. 
Section~\ref{openQ}
introduces modeling of open quantum systems and the computation of their
states while accounting for dynamics and interactions with the
environment. Section~\ref{software} describes the software architecture, while
Section~\ref{algo} presents the PQLS used algorithm. Example applications of
the library are provided in Section~\ref{example}, and performance benchmarks
are presented in Section~\ref{sec:benchmarks}. Finally,
Section~\ref{conclusion} concludes the paper.


\section{Modeling Open Quantum Systems}
\label{openQ}

Realistic atomic systems are not  perfectly
isolated from their surroundings. Instead, they interact with
environmental degrees of freedom, forming open quantum systems
\cite{agarwal2012open}. 
In general, an open quantum system can be described as a quantum
system $S$, associated with a Hilbert space $\mathcal{H}_S$, that is
coupled to another quantum system $B$, representing its environment
and associated with a Hilbert space $\mathcal{H}_B$. Therefore, the
system $S$ can be regarded as a subsystem of the combined system
$S+B$, whose Hilbert space is given by the tensor-product space
$\mathcal{H}_S \otimes \mathcal{H}_B$~\cite{openq}.

The interaction with the environment introduces dissipative effects into the
dynamics of the atomic system. The temporal evolution of the atomic density
matrix $\boldsymbol{\rho}$ can be described by the Lindblad master
equation~\cite{eq1source,linblad2}:
\begin{equation}
\frac{d\boldsymbol{\rho}}{dt}
=
-\frac{i}{\hbar}
[\boldsymbol{H},\boldsymbol{\rho}]
+
\mathcal{D}(\boldsymbol{\rho})
\equiv
\mathcal{L}(\boldsymbol{\rho}),
\label{lind}
\end{equation}
where $\boldsymbol{H}$ is the Hamiltonian,
$\mathcal{D}(\boldsymbol{\rho})$ represents the dissipative contribution due
to coupling with the environment, and
$[\boldsymbol{H},\boldsymbol{\rho}]
=\boldsymbol{H}\boldsymbol{\rho}
-\boldsymbol{\rho}\boldsymbol{H}$ is the commutator.
The Liouvillian superoperator $\mathcal{L}$ therefore combines the coherent
Hamiltonian dynamics and dissipative processes into a single representation
of the system dynamics~\cite{Liouvillan}.

For a coherently driven $N$-level system, the Hamiltonian under the
rotating-wave approximation can be represented as an $N\times N$
Hermitian matrix, with effective detunings on the diagonal and Rabi
frequencies describing the couplings between states on the
off-diagonal elements~\cite{shore2008coherent}:
\begin{equation}
\boldsymbol{H}
=
\frac{\hbar}{2}
\begin{bmatrix}
2\Delta_1
    & \Omega_{12}
    & \Omega_{13}
    & \cdots
    & \Omega_{1N}
\\
\Omega_{12}^{*}
    & 2\Delta_2
    & \Omega_{23}
    & \cdots
    & \Omega_{2N}
\\
\Omega_{13}^{*}
    & \Omega_{23}^{*}
    & 2\Delta_3
    & \cdots
    & \Omega_{3N}
\\
\vdots
    & \vdots
    & \vdots
    & \ddots
    & \vdots
\\
\Omega_{1N}^{*}
    & \Omega_{2N}^{*}
    & \Omega_{3N}^{*}
    & \cdots
    & 2\Delta_N
\end{bmatrix},
\label{general_hamiltonian}
\end{equation}
where $\Delta_i$ denotes the effective detuning of state $|i\rangle$
and $\Omega_{ij}$ is the Rabi frequency coupling states $|i\rangle$
and $|j\rangle$. For states that are not coherently coupled,
$\Omega_{ij}=0$.

The dissipative contribution is described by the Lindblad dissipator~\cite{decay},
\begin{equation}
\label{L}
\mathcal{D}(\boldsymbol{\rho})
=
\sum_k
\left(
\boldsymbol{C}_k \boldsymbol{\rho} \boldsymbol{C}_k^\dagger
-
\frac{1}{2}
\left\{
\boldsymbol{C}_k^\dagger \boldsymbol{C}_k,
\boldsymbol{\rho}
\right\}
\right),
\end{equation}
where $\boldsymbol{C}_k$ denotes the collapse operator associated with
the $k$th dissipative channel.
For a decay from state $|i\rangle$ to $|j\rangle$ with rate
$\Gamma_{ij}$, $\boldsymbol{C}_{ij}
=\sqrt{\Gamma_{ij}}\,|j\rangle\langle i|$.
The decay rate $\Gamma_{ij}$ may include any incoherent process that
transfers population from state $|i\rangle$ to state $|j\rangle$,
such as spontaneous emission, blackbody-radiation-induced transitions,
collisional relaxation, or other population-loss mechanisms~\cite{farley1981blackbody}.


Open quantum-system models provide a physically meaningful description of
quantum systems by accounting for their interaction with the surrounding
environment. As introduced earlier, Rydberg atoms are highly sensitive to external
electromagnetic fields, making their modeling an important application of
open quantum-system theory.~\cite{GB2025}. In this framework, coherent interactions induced by
external fields are represented by the Hamiltonian
$\boldsymbol{H}$, while dissipative processes such as spontaneous
emission, blackbody-radiation-induced transitions, and other relaxation
mechanisms are incorporated through the collapse operators in
$\mathcal{D}(\boldsymbol{\rho})$.


\begin{figure*}[!t]
    \centering
    \includegraphics[width=\textwidth]{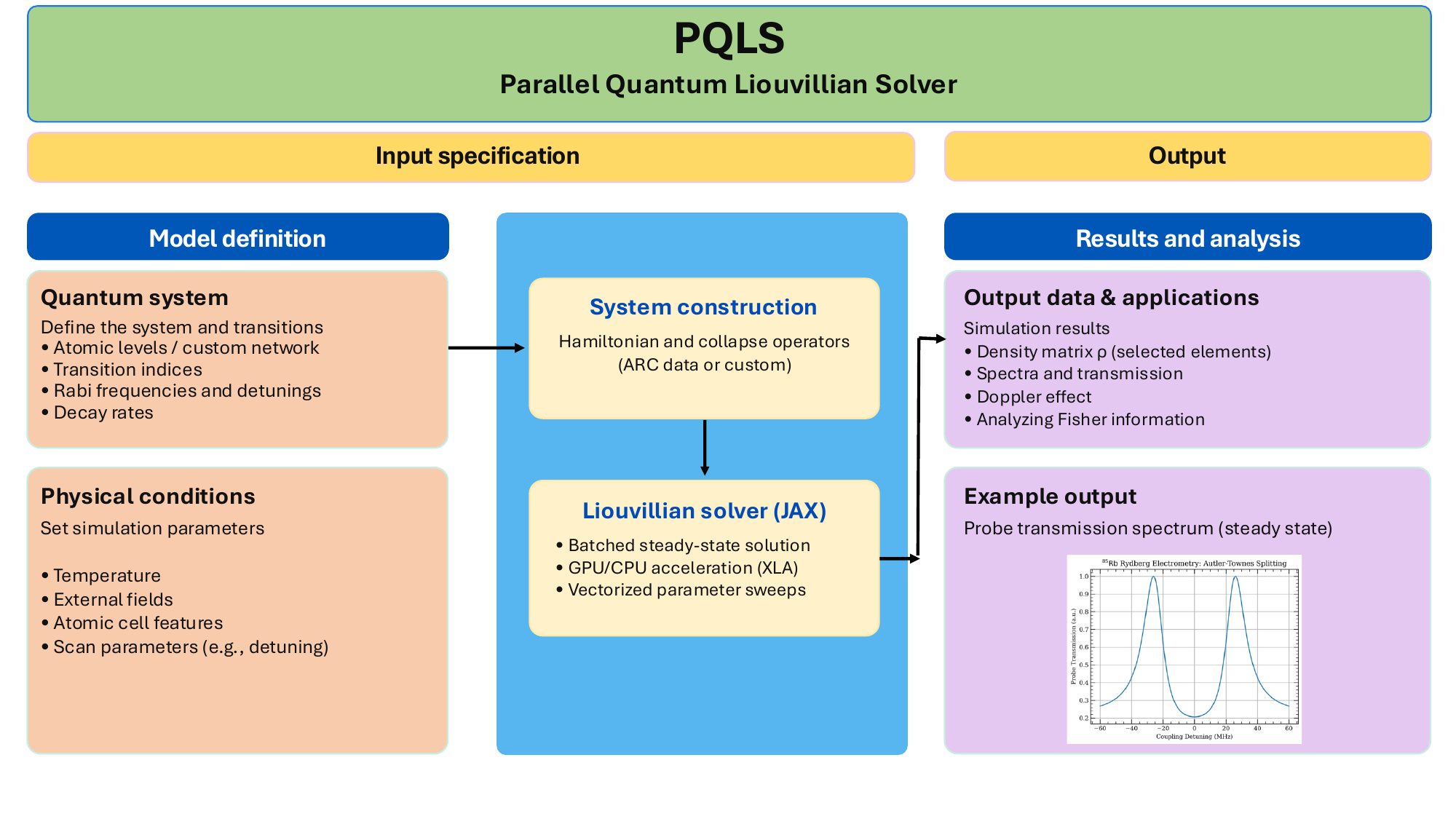}
    \caption{PQLS algorithm at a glance.}
    \label{fig:pqls}
\end{figure*}

\section{Software Design}
\label{software}
PQLS is specifically designed to minimize the computational overhead typically associated with computing the steady-state of the Lindblad master equation. By leveraging JAX kernels optimized via the XLA (Accelerated Linear Algebra) compiler~\cite{jax2018github}, PQLS significantly outperforms traditional NumPy~\cite{harris2020array} or iterative QuTiP-based implementations~\cite{qutip5}, while providing new capabilities for researchers. An overview of PQLS algorithm is represented in Fig.~\ref{fig:pqls}, explained in detail in the upcoming sections.

\subsection{Physics-Aware Public API}
\begin{figure*}[!t]
    \centering
    \includegraphics[width=\textwidth]{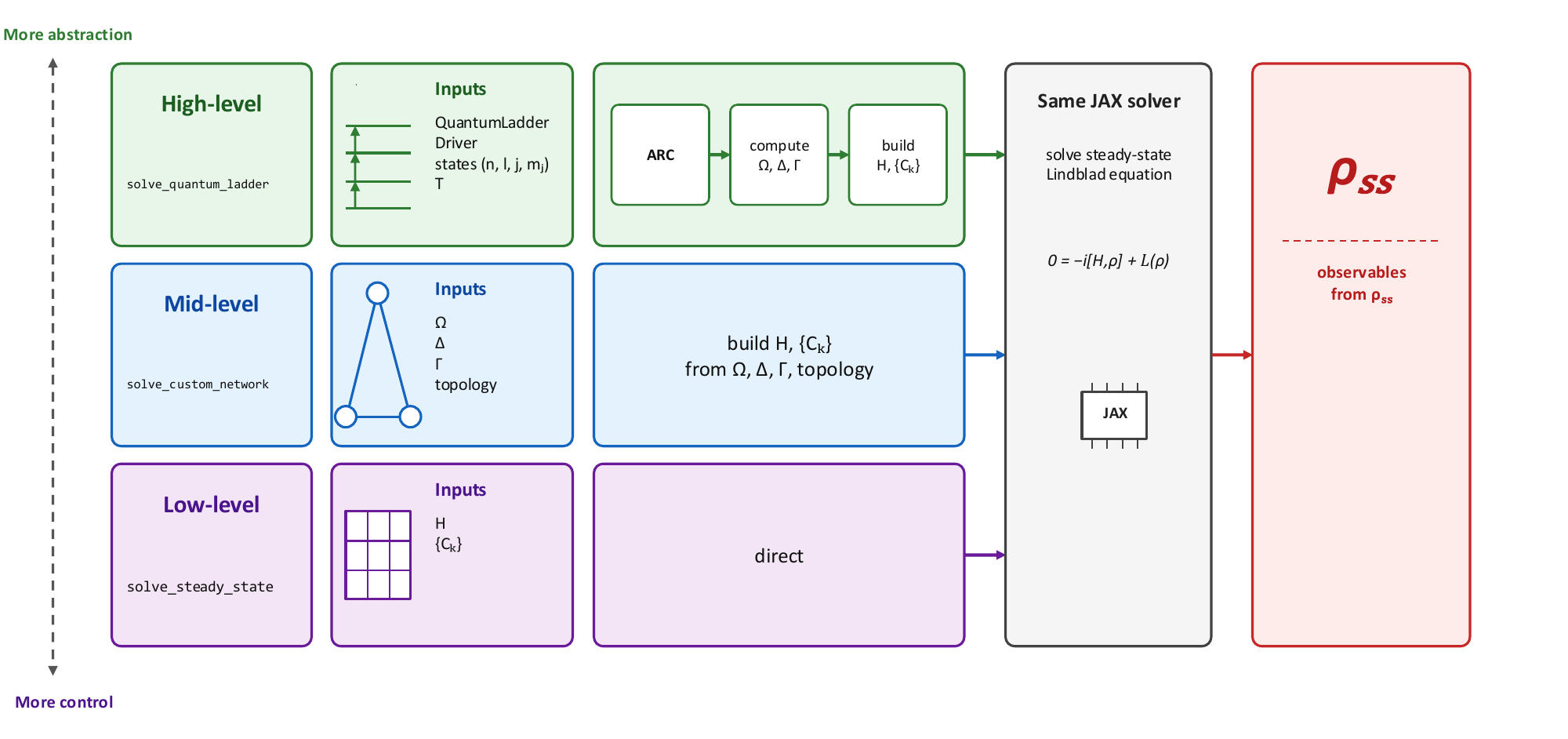}
    \caption{Three-tier user-facing API of PQLS.}
    \label{fig:pqls_tiers}
\end{figure*}

To accommodate diverse simulation requirements, the user-facing API is structured into three distinct tiers of abstraction, represented in Fig.~\ref{fig:pqls_tiers}, at the top of the next page.

At the highest level, \texttt{solve\_quantum\_ladder} provides domain-specific support for $N$-level atomic ladder systems. Users define atomic topologies using physics-oriented configuration objects,
such as the \texttt{QuantumLadder} and \texttt{Driver} classes, with each
energy level specified by the quantum numbers $(n,l,j,m_j)$, where $n$, $l$,
$j$, and $m_j$ denote the principal, orbital angular momentum, total angular
momentum, and magnetic quantum numbers, respectively. The high-level API natively supports Rubidium-85 ($^{85}\text{Rb}$), Rubidium-87 ($^{87}\text{Rb}$), and Cesium (Cs) atoms. The solver also requires the ambient temperature, $T$, as an input to accurately compute temperature-dependent blackbody-radiation-induced transitions. 

When utilizing \texttt{solve\_custom\_network}, the intermediate  API, researchers explicitly define system topologies using transition tuples (e.g., $(i, j)$ connecting lower state $i$ to upper state $j$) and decay tuples (e.g., $(j, i, \Gamma)$ denoting a spontaneous decay from state $j$ to state $i$ at rate $\Gamma$). These parameters are used to construct the collapse operators and are mapped to the Lindblad superoperator using the Kronecker product structures detailed in Appendix \ref{app:liouvillian}. 

Rather than requiring researchers to manually calculate matrices~\eqref{general_hamiltonian} and~\eqref{L} for ladder topologies, the top-level API automatically queries the ARC database to resolve fundamental physical properties. During this process, the high-level API automatically computes the transition
dipole matrix elements, which, together with the applied driving-field
amplitudes and polarizations, determine the corresponding Rabi frequencies.
It also computes the resonant transition frequencies and
temperature-dependent decay rates, including contributions from spontaneous
emission and blackbody-radiation-induced transitions. If users wish to include
additional decay or dephasing mechanisms beyond those provided automatically,
they may instead use the low-level API, where the Hamiltonian and collapse
operators can be specified directly. 

Because querying external atomic databases introduces initialization overhead, PQLS exposes intermediate and low-level APIs for pure numerical execution. The mid-level \texttt{solve\_custom\_network} function bypasses ARC entirely, enabling researchers to simulate arbitrary network topologies (e.g., $\Lambda$-systems with branching decays) using raw Rabi frequencies and detunings. Finally, the low-level \texttt{solve\_steady\_state} API allows advanced users to directly pass batched Hamiltonian tensors and collapse (jump) operators into the JAX solver.

\subsection{JAX Backend}
Beneath the physics abstraction layer, the core execution engine of PQLS is implemented entirely in JAX. The solver maps the parameters retrieved from the frontend into tensors representing the system's Hamiltonian and collapse operators.

The underlying JAX architecture supports hardware-independent execution,
allowing the same PQLS code to run on CPUs or supported GPUs. As a result,
large batches of steady-state problems can be executed in parallel on GPU
hardware without requiring changes to the user-level simulation code.

Furthermore, because JAX provides native automatic differentiation,
the solver outputs can be differentiated with respect to model parameters
with minimal additional code. This native differentiability provides a significant advantage for advanced quantum sensing and communication applications, such as calculating Fisher information or performing gradient-based optimization of driving fields.

\section{PQLS Implementation}
\label{algo}
To efficiently compute the steady-state of a quantum system, PQLS translates the physical parameters into a vectorized linear algebra problem. 
The algorithm is based on the Lindblad formulation of open quantum systems and reformulates the dynamics in terms of the Liouvillian superoperator to obtain the steady-state density matrix.

\subsection{Superoperator Computation}
To find the steady-state condition ($\dot{\boldsymbol{\rho}} = 0$), PQLS maps the Lindblad equation into Liouville space. By vectorizing the $N \times N$ density matrix $\boldsymbol{\rho}$ into a column vector $\vec{\boldsymbol{\rho}}$ of size $N^2$, the master equation can be rewritten as a linear system:
\begin{equation}
    \mathcal{L} \vec{\boldsymbol{\rho}} = 0
\end{equation}
where $\mathcal{L}$ is the Liouvillian superoperator. Its construction from
the Hamiltonian and collapse operators using Kronecker products is described
in Appendix.

Because the steady-state Liouvillian is singular, the system of equations is underdetermined. PQLS enforces the physical probability conservation constraint, $\text{Tr}(\boldsymbol{\rho}) = 1$, by replacing the first row of $\mathcal{L}$ with the trace condition, and setting the corresponding first element of the target vector $\vec{\boldsymbol{b}}$ to 1. This transforms the problem into a non-singular linear system, $\mathcal{L}' \vec{\boldsymbol{\rho}} = \vec{\boldsymbol{b}}$, which is solved via JAX's optimized \texttt{jnp.linalg.solve} backend with LU factorization.

\subsection{Hardware-Accelerated Vectorization}
The primary algorithmic advantage of PQLS lies in its approach to parameter sweeps. Traditional solvers typically iterate over independent parameter values
(e.g., a sweep of $B$ laser detunings, where $B$ denotes the batch size)
using \texttt{for} loops, invoking the Lindblad solver $B$ separate times.

Instead, PQLS uses JAX's vectorized map transformation
(\texttt{jax.vmap}) to process the full batch collectively.
For a batch size of $B$, the Hamiltonians are represented as a
single three-dimensional tensor of shape $(B,N,N)$, where each
$N\times N$ slice corresponds to one independent parameter configuration. The \texttt{vmap} primitive pushes the iterative loop down to the XLA compiler level, which optimally chunks the $B$ independent linear systems across the available hardware cores (CPU or GPU). This allows PQLS to solve thousands of steady-states simultaneously in a single compiled execution step, drastically reducing the Python interpreter overhead and optimizing hardware utilization.

\section{Validation and Accuracy}
\label{valid}
Because PQLS bypasses standard sequential solvers in favor of a custom
JAX-compiled steady-state kernel, its numerical accuracy was validated against
established frameworks to verify the correctness of the implementation. To verify numerical consistency, PQLS includes a rigorous Continuous Integration (CI) test suite that utilizes \texttt{pytest} parameterized testing to validate the JAX backend against  RydIQule and QuTiP's \texttt{steadystate} solvers across all API abstraction tiers.

\subsection{Multi-Tiered API Validation}
To ensure robustness, the CI suite evaluates the solver through highly specific physical regimes as well as arbitrary statistical matrices:

\begin{itemize}
    \item \textbf{Physics-Aware End-to-End Testing:} The top-level API is validated by simulating a 4-level $^{85}\text{Rb}$ Rydberg RF ladder ($5S_{1/2} \to 5P_{3/2} \to 50D_{5/2} \to 51P_{3/2}$). This ensures the solver accurately integrates with the ARC database to derive transition dipole matrix elements, temperature-dependent decay pathways, and correctly handles both amplitude and detuning parameter sweeps.
    \item \textbf{Custom Network Topologies:} The mid-level API is tested using 3-level systems with branching decay pathways. The test suite verifies batched parameter sweeps by comparing each PQLS
solution with the corresponding reference solution obtained from QuTiP
using sequential steady-state solves.

    \item \textbf{Randomized Hamiltonian Testing:} To decouple validation from specific atomic physics, the low-level API dynamically generates randomized, Hermitian Hamiltonians for arbitrary $N$-level systems ($N \in \{2, 3, 4, 5, 6\}$) alongside sets of cascade jump operators between random pairs of levels. These are fed into PQLS across varying batch sizes (from unbatched scalars up to 128-dimensional parameter sweeps) to verify the JAX tensor broadcasting logic.
\end{itemize}

Beyond agreement with external reference solvers, PQLS also verifies that the
computed density matrices, $\boldsymbol{\rho}$, satisfy the fundamental
physical requirements of a valid quantum state. To distinguish true violations
from floating-point round-off errors, the test suite defines consistent
numerical tolerances for these checks:
\begin{itemize}
    \item \textbf{Parity Tolerance ($\mathbf{10^{-11}}$):} The maximum allowable absolute difference between the PQLS JAX backend and QuTiP's ground truth across randomized trials.
    \item \textbf{Invariant Tolerance ($\mathbf{10^{-14}}$):} The strict precision threshold applied to fundamental physical invariants. The suite explicitly asserts Probability Conservation ($\text{Tr}(\boldsymbol{\rho}) = 1$) and Observability/Hermiticity ($\boldsymbol{\rho} = \boldsymbol{\rho}^\dagger$) to prevent the solver from leaking population states.
   \item \textbf{Positive Semi-Definite Noise Floor ($\mathbf{-10^{-13}}$):}
The eigenvalues $\lambda_i$ of the density matrix $\boldsymbol{\rho}$,
defined through
$\boldsymbol{\rho}\mathbf{v}_i=\lambda_i\mathbf{v}_i$,
are required to satisfy $\lambda_i \ge -10^{-13}$.
For a physically valid density matrix, these eigenvalues must be
nonnegative because they represent the probabilities associated with the
states in the spectral decomposition of $\boldsymbol{\rho}$.
The small negative threshold allows for floating-point round-off errors.
  \item \textbf{Minimum Coherence Threshold ($10^{-6}$):}
For validation cases involving coherent optical response, such as the Rydberg
EIT/ATS benchmarks, the test suite requires at least one off-diagonal
density-matrix element to satisfy
$\lvert \rho_{ij} \rvert > 10^{-6}$ for $i\neq j$.
This ensures that the validation includes cases with meaningful off-diagonal
coherences, which are central to the optical response in Rydberg EIT/ATS
simulations.
\end{itemize}

\section{Simulation Examples}
\label{example}
To demonstrate the physical accuracy and API flexibility of PQLS, the package includes several representative quantum optical phenomena as built-in examples. Here we report on some key results.

\subsection{AT Splitting}
To demonstrate the flexibility of the PQLS architecture, we evaluate a 4-level $^{85}\text{Rb}$ Rydberg electrometry system ($5S_{1/2} \to 5P_{3/2} \to 50D_{5/2} \to 51P_{3/2}$) using both the highest and lowest tiers of the API. 

Using the high-level \texttt{solve\_quantum\_ladder} API, PQLS natively handles the decay pathways and calculates probe transmission while sweeping the coupling field detuning. Alternatively, bypassing the physical abstractions, the low-level \texttt{solve\_steady\_state} API allows users to manually construct and input the batched Hamiltonian tensor of shape $(B, N, N)$ alongside the spontaneous decay jump operators. 

\begin{figure}[htbp]
\centerline{\includegraphics[width=\columnwidth]{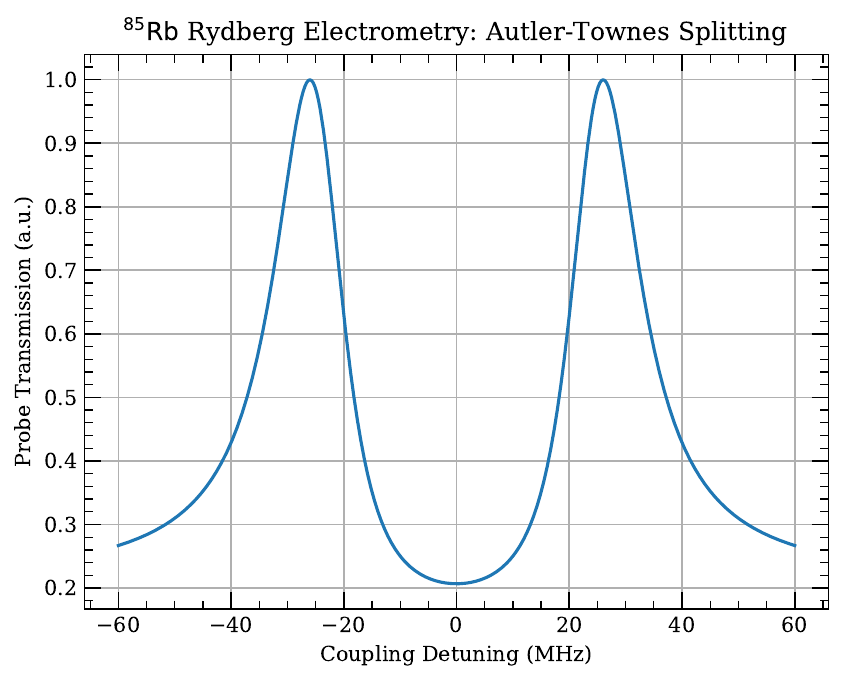}}
\caption{Probe transmission spectrum of a 4-level $^{85}\text{Rb}$ ladder system exhibiting ATS. This identical chart is generated whether using the high-level \texttt{solve\_quantum\_ladder} API or by directly passing batched Hamiltonian tensors into the \texttt{solve\_steady\_state} API.}
\label{fig:ats}
\end{figure}

Both approaches yield the exact same numerical spectrum, successfully reproducing the signature ATS peaks as shown in Fig.~\ref{fig:ats}.

\subsection{Coherent Population Trapping (CPT)}

To demonstrate \texttt{solve\_custom\_network}, the intermediate  API, we
consider a three-level $\Lambda$-system in $^{87}\mathrm{Rb}$, where two
lower states are optically coupled to a common excited state. The model
includes branching spontaneous-emission channels from the excited state to
both lower states, demonstrating the ability of the mid-level API to represent
non-ladder network topologies.

\begin{figure}[htbp]
\centerline{\includegraphics[width=\columnwidth]{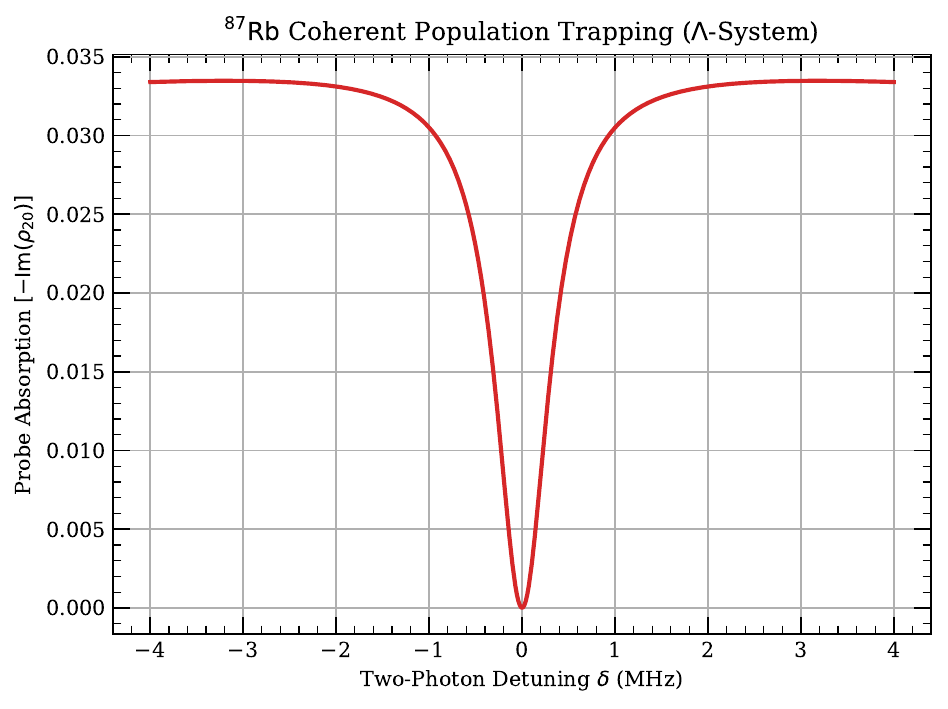}}
\caption{Probe absorption spectrum in a 3-level $^{87}\text{Rb}$ $\Lambda$-system.}
\label{fig:cpt}
\end{figure}

Fig.~\ref{fig:cpt} shows the probe absorption as a function of the
two-photon detuning, defined as the mismatch between the frequency difference
of the two driving fields and the energy splitting between the two lower
states. At zero two-photon detuning, destructive interference between the excitation
pathways forms a dark state, suppressing probe absorption and producing a
transparency window. Away from resonance, the dark-state condition is lost
and the absorption increases.

\begin{figure*}[!t]
\centerline{\includegraphics[width=\textwidth]{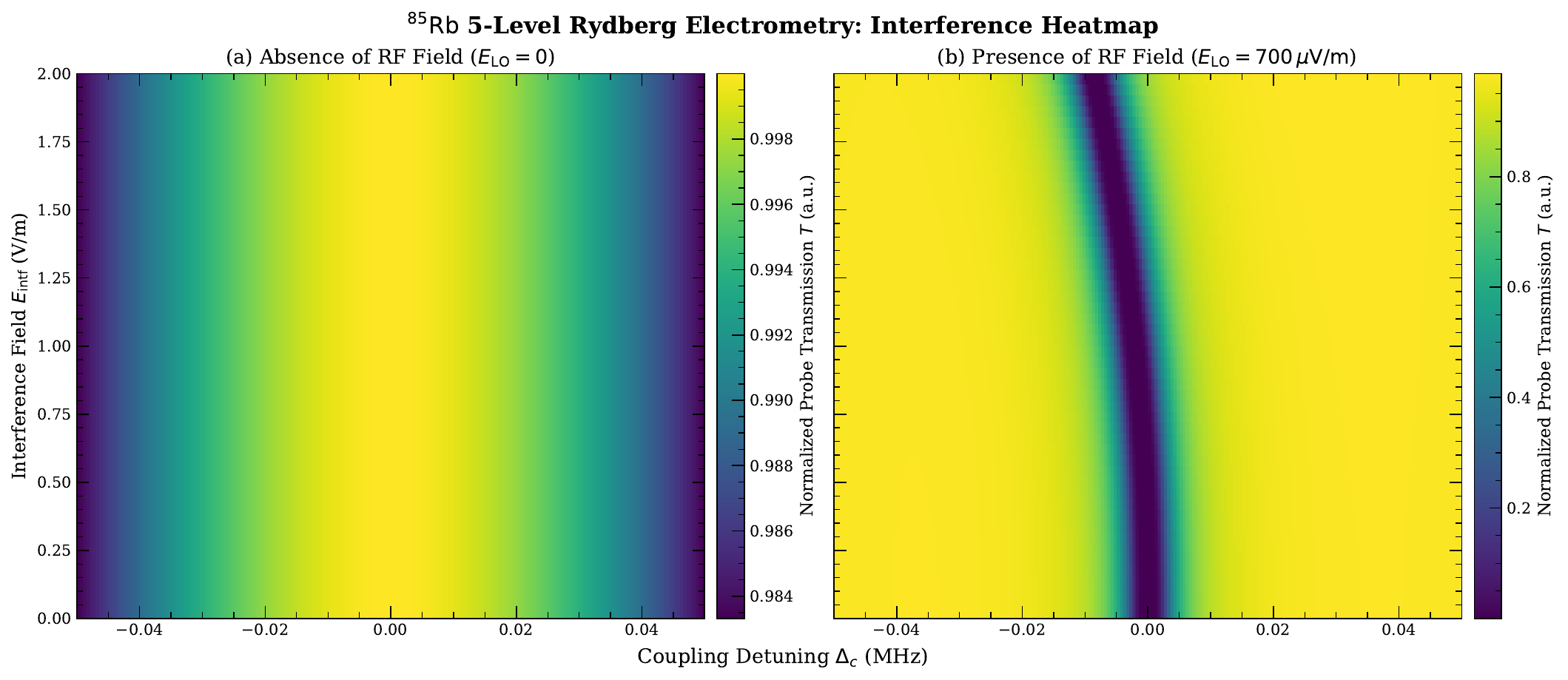}}
\caption{2D transmission heatmaps reproducing the interference study from \cite{GB2025}, generated by evaluating 10,000 independent steady-state configurations of a 5-level $^{85}\text{Rb}$ system. PQLS computes this entire 2D parameter space natively on GPU hardware in approximately 10 milliseconds.}
\label{fig:heatmap}
\end{figure*}
\subsection{2D Parameter Sweeps and Heatmaps}
To highlight the advantages of XLA-accelerated vectorization, we ported the 5-level $^{85}\text{Rb}$ Rydberg receiver simulation originally presented by Rostampoor and Adve \cite{GB2025} into the PQLS framework using the mid-level \texttt{solve\_ladder\_system} API. The original study investigated the effects of interference by computing transmission heatmaps across a dense 2D parameter grid (simultaneously varying both the interference field amplitude and the coupling laser detuning). To showcase the execution speed of our solver, PQLS completes 10,000 independent steady-state computations to perfectly reproduce these heatmaps, as shown in Fig. \ref{fig:heatmap}, at the top of the next page. Deployed on the GPU hardware detailed in Section \ref{sec:benchmarks}, PQLS evaluates this 
entire 2D parameter space in approximately 10 milliseconds.

\section{Benchmarks}
\label{sec:benchmarks}

To rigorously evaluate the computational efficiency of PQLS, we benchmarked its execution time against three existing frameworks: traditional QuTiP, QuTiP-JAX, and RydIQule. All benchmarks simulated the steady-state of a 4-level $^{87}\text{Rb}$  system across varying parameter space sizes. 

All performance metrics were recorded on a desktop workstation operating a Linux environment, equipped with an AMD Ryzen 9 7950X 16-Core processor, 32 GB of system RAM, and an NVIDIA GeForce RTX 4090 GPU (24 GB VRAM).

\subsection{Comparison with QuTiP and QuTiP-JAX}
Standard implementations of the Lindblad master equation typically rely on sequential evaluation. Table~\ref{tab:qutip_bench} presents the wall-clock
execution time, i.e., the total elapsed time from the start to completion of
the computation, required to solve a 1,000-point parameter sweep
(e.g., varying the coupling-laser detuning).

While \texttt{qutip-jax} brings hardware acceleration to traditional solvers
by replacing the underlying matrix representations with JAX tensors, the
parameter-sweep workflow still relies on Python-level control flow to construct
individual Liouvillians. This limits compiler-level vectorization across the
sweep using \texttt{jax.vmap} and introduces substantial XLA dispatch overhead. \footnote{
XLA dispatch overhead refers to the runtime cost of repeatedly launching
individual compiled operations from the host program to the execution backend.}

\begin{figure*}[htbp]
\centerline{\includegraphics[width=\textwidth]{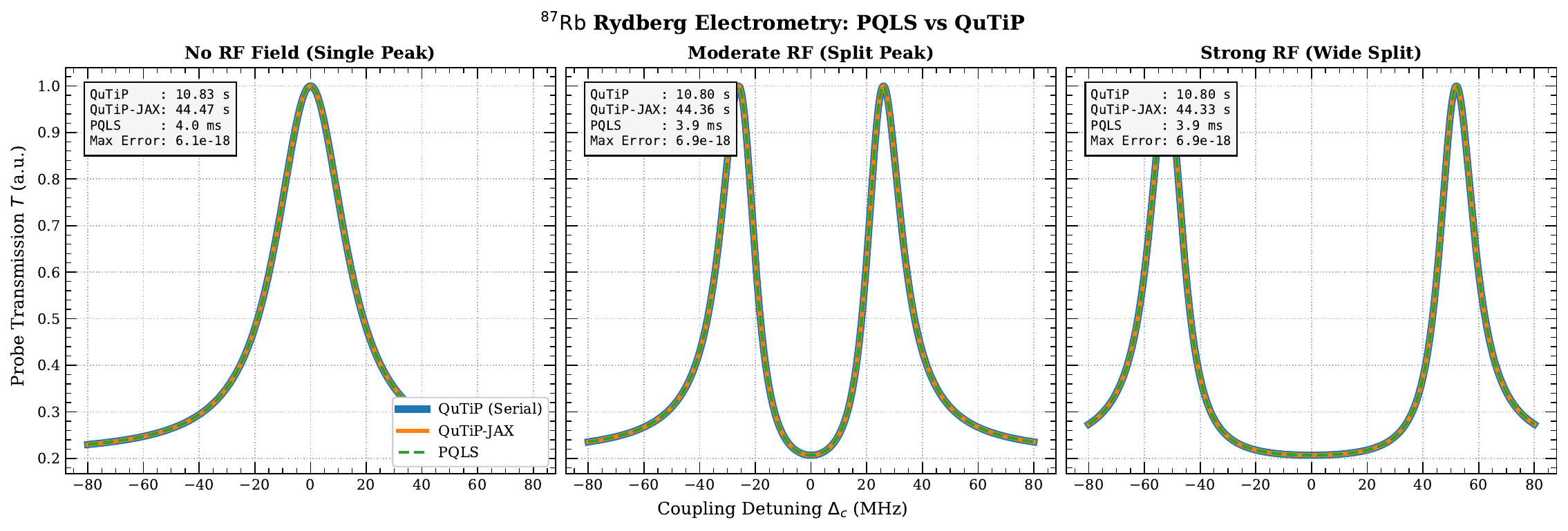}}
\caption{Comparison of probe transmission spectra for a 4-level $^{87}\text{Rb}$ Rydberg electrometry system under varying RF field strengths ($0.0$, $2.0$, and $4.0$ V/m). The charts show a 1,000-point parameter sweep of the coupling laser detuning. The precisely overlaid curves demonstrate numerical parity between PQLS (dashed green) and the reference QuTiP implementations (solid lines). Inset boxes detail the maximum absolute error alongside wall-clock execution times (averaged over 10 trials), highlighting the orders-of-magnitude computational speedup achieved by PQLS.}
\label{fig:qutip_bench}
\end{figure*}

By contrast, PQLS represents the entire 1,000-point parameter sweep as a
single batched computation compiled by XLA. As shown in
Table~\ref{tab:qutip_bench}, this approach yields a speedup of more than
$2{,}700\times$ over both the standard QuTiP and QuTiP-JAX implementations,
while maintaining agreement with the reference solutions within the specified
numerical tolerance. Fig.~\ref{fig:qutip_bench} visually confirms this exact numerical parity across varying RF field strengths, demonstrating that PQLS perfectly reproduces QuTiP's transmission spectra while operating orders of magnitude faster. Throughput (sys/s) denotes the number of independent steady-state systems solved per second.

\begin{table}[htbp]
\caption{Execution Time for a 1,000-Point Parameter Sweep}
\begin{center}
\resizebox{\columnwidth}{!}{
\begin{tabular}{lccc}
\toprule
\textbf{Framework} & \textbf{Time (s)} & \textbf{Throughput (sys/s)} & \textbf{Speedup} \\
\midrule
QuTiP (Serial) & 10.810 & 92 & $1\times$ \\
QuTiP-JAX & 44.385 & 22 & $0.24\times$ \\
\textbf{PQLS (Ours)} & \textbf{0.0039} & \textbf{253,787} & \textbf{$\sim$2,743$\times$} \\
\bottomrule
\end{tabular}
}
\label{tab:qutip_bench}
\end{center}
\end{table}

\subsection{Scaling Performance vs. RydIQule}
As a final evaluation, we compared the scalability of PQLS against
RydIQule, a specialized Rydberg solver that mitigates Python overhead
via NumPy array stacking. 

\begin{figure}[htbp]
\centerline{\includegraphics[width=\columnwidth]{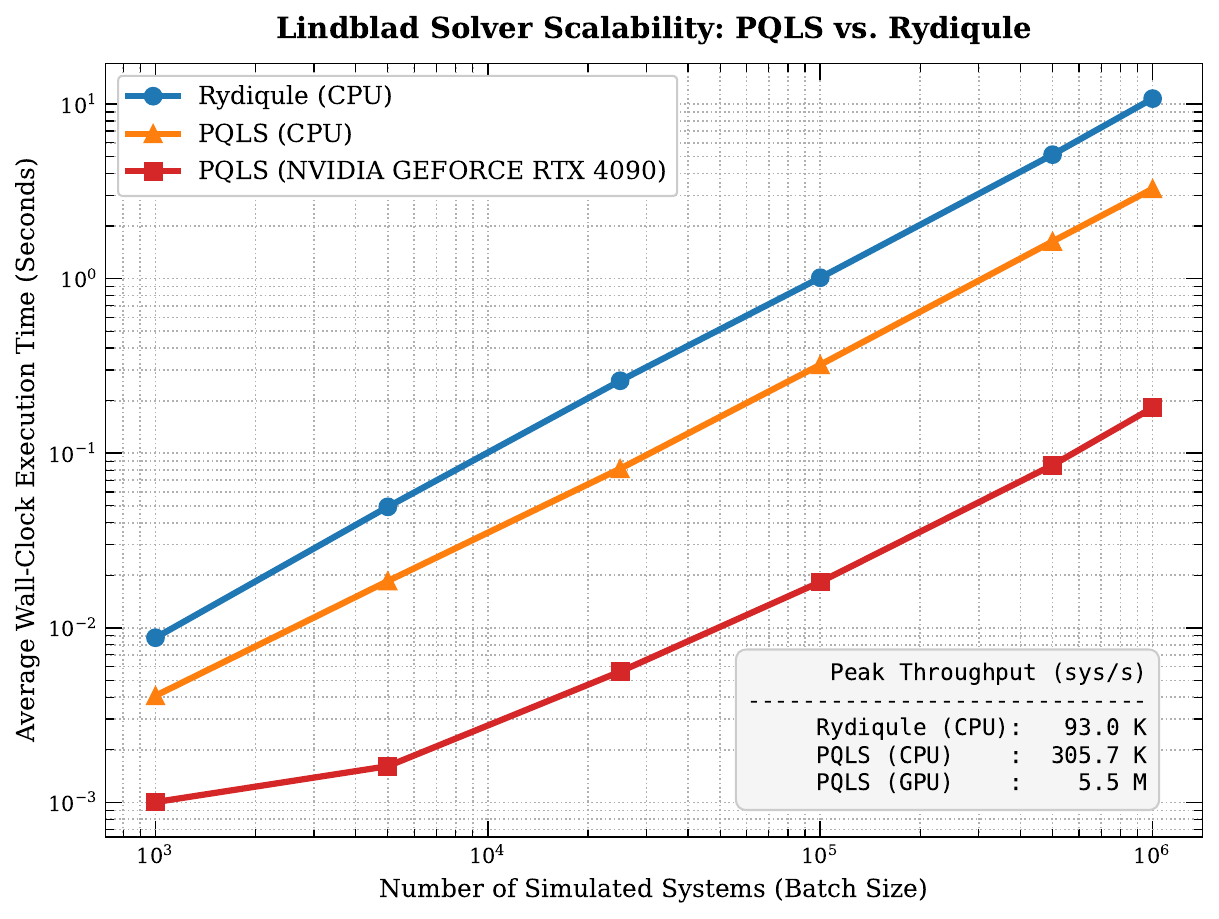}}
\caption{Log-log scalability comparison demonstrating the wall-clock execution time of PQLS versus RydIQule across increasing batch sizes.}
\label{fig:scaling}
\end{figure}

While NumPy vectorization is highly efficient for small batch sizes, it is fundamentally bottlenecked by CPU execution speed and system RAM bandwidth. Table \ref{tab:scaling} demonstrates the scaling laws of both solvers from $10^3$ to $10^6$ parameters. At $1,000,000$ points, RydIQule requires over 10 seconds to complete the calculation. Executing PQLS on the exact same CPU completes the sweep in 3.27 seconds, demonstrating the algorithmic superiority of XLA compilation over NumPy stacking. When PQLS is deployed on GPU hardware (NVIDIA RTX 4090), it natively saturates the memory bandwidth to finish the identical parameter sweep in less than 0.2 seconds. This yields a $58\times$ wall-clock speedup over the CPU-bound baseline, achieving a peak throughput of over $5.4$ million (sys/s). Fig.~\ref{fig:scaling} compares PQLS and RydIQule in terms of average
wall-clock execution time as a function of the number of simulated systems.
The results show that PQLS consistently outperforms RydIQule, with the highest
performance achieved using GPU acceleration.

\begin{table}[htbp]
\caption{Scalability Benchmark: PQLS vs. RydIQule}
\begin{center}
\resizebox{\columnwidth}{!}{
\begin{tabular}{rcccc}
\toprule
\textbf{Batch ($N$)} & \textbf{Ryd (CPU s)} & \textbf{PQLS (CPU s)} & \textbf{PQLS (GPU s)} & \textbf{GPU Speedup} \\
\midrule
1,000 & 0.009 & 0.004 & 0.001 & $8\times$ \\
25,000 & 0.260 & 0.081 & 0.006 & $46\times$ \\
100,000 & 1.012 & 0.321 & 0.018 & $55\times$ \\
1,000,000 & 10.749 & 3.271 & 0.182 & $58\times$ \\
\bottomrule
\end{tabular}
}
\label{tab:scaling}
\end{center}
\end{table}

\section{Conclusions}
\label{conclusion}
In this paper, we introduced PQLS, a Python library designed to improve the
efficiency of steady-state simulations of open quantum systems, particularly
for computationally intensive parameter sweeps. PQLS exploits vectorization to
reduce Python-level overhead by processing batches of system configurations
simultaneously and uses JAX to accelerate the solution of the Lindblad master
equation. To accommodate different modeling requirements, the library provides
three levels of API abstraction, ranging from a low-level interface that offers
maximum control over the numerical model to a high-level physics-oriented
interface that automatically performs the calculations required to construct
and solve the corresponding Liouvillian problem. The APIs were validated and
benchmarked against existing Python packages, including QuTiP, QuTiP-JAX, and
RydIQule, in terms of both numerical accuracy and computational performance.
The simulation results and benchmarks demonstrate substantial performance
gains for PQLS. For a 1,000-point parameter sweep, PQLS achieved a throughput
of 253,787 systems/s and an approximately $2{,}743\times$ speedup over serial
QuTiP. For one million simulated systems, PQLS reduced the CPU execution time
from 10.749~s with RydIQule to 3.271~s, while GPU acceleration further reduced
it to 0.182~s, corresponding to a reported speedup of up to $58\times$.

\appendix[General Construction of the Steady-State Liouvillian]
\label{app:liouvillian}

Consider a general $N$-level open quantum system described by a Hamiltonian
\begin{equation}
    \boldsymbol{H} \in \mathbb{C}^{N\times N}.
\end{equation}
The density matrix $\boldsymbol{\rho} \in \mathbb{C}^{N\times N}$ evolves according to the
Lindblad master equation
\begin{equation}
    \frac{d\boldsymbol{\rho}}{dt}
    =
    -i[\boldsymbol{H},\boldsymbol{\rho}]
    +
    \sum_{k=1}^{M}
    \left(
        \boldsymbol{C}_k \boldsymbol{\rho} \boldsymbol{C}_k^\dagger
        -
        \frac{1}{2} \boldsymbol{C}_k^\dagger \boldsymbol{C}_k \boldsymbol{\rho}
        -
        \frac{1}{2} \boldsymbol{\rho} \boldsymbol{C}_k^\dagger \boldsymbol{C}_k
    \right),
    \label{eq:lindblad_general}
\end{equation}
where $M$ is the number of decay channels and $\boldsymbol{C}_k$ is the collapse operator
associated with the $k$th channel.

For a decay from state $\lvert s_k\rangle$ to state $\lvert d_k\rangle$
with rate $\gamma_k$, the corresponding collapse operator is
\begin{equation}
    \boldsymbol{C}_k
    =
    \sqrt{\gamma_k}
    \lvert d_k\rangle\langle s_k\rvert,
    \qquad k=1,\ldots,M,
    \label{eq:collapse_general}
\end{equation}
where $\gamma_k$ denotes the decay rate of the corresponding transition.

To express Eq.~\eqref{eq:lindblad_general} as a linear system, the density
matrix is vectorized as
\begin{equation}
    \mathbf{r}
    =
    \operatorname{vec}(\boldsymbol{\rho})
    \in \mathbb{C}^{N^2}.
\end{equation}
Using row-wise vectorization,
\begin{equation}
    \operatorname{vec}(\boldsymbol{A}\boldsymbol{X}\boldsymbol{B})
    =
    (\boldsymbol{A}\otimes \boldsymbol{B}^{T})\operatorname{vec}(\boldsymbol{X}),
\end{equation}
where $\otimes$ denotes the Kronecker product and $\boldsymbol{I}_N$ denotes the
$N\times N$ identity matrix.

The coherent contribution to the Liouvillian is
\begin{equation}
    \boldsymbol{\mathcal{L}}_{\mathrm{H}}
    =
    -i
    \left(
        \boldsymbol{H}\otimes \boldsymbol{I}_N
        -
        \boldsymbol{I}_N\otimes \boldsymbol{H}^{T}
    \right).
    \label{eq:L_H_general}
\end{equation}

The dissipative contribution is
\begin{equation}
\begin{aligned}
    \boldsymbol{\mathcal{L}}_{\mathrm{D}}
    =
    \sum_{k=1}^{M}
    \Bigg[
        &\boldsymbol{C}_k\otimes \boldsymbol{C}_k^{*}
        -
        \frac{1}{2}
        \left(
            \boldsymbol{C}_k^\dagger \boldsymbol{C}_k\otimes \boldsymbol{I}_N
        \right)-
        \frac{1}{2}
        \left(
            \boldsymbol{I}_N\otimes \boldsymbol{C}_k^{T}\boldsymbol{C}_k^{*}
        \right)
    \Bigg].
\end{aligned}
\label{eq:L_D_general}
\end{equation}

Here, $(\cdot)^T$, $(\cdot)^*$, and $(\cdot)^\dagger$ denote the transpose,
complex conjugate, and Hermitian conjugate, respectively.

The complete Liouvillian is therefore
\begin{equation}
    \boldsymbol{\mathcal{L}}
    =
    \boldsymbol{\mathcal{L}}_{\mathrm{H}}
    +
    \boldsymbol{\mathcal{L}}_{\mathrm{D}},
    \qquad
    \boldsymbol{\mathcal{L}}\in\mathbb{C}^{N^2\times N^2}.
    \label{eq:L_general}
\end{equation}

	\bibliography{Bib}
	\bibliographystyle{IEEEtran}

\end{document}